# YOUNG STAR CLUSTERS IN THE CIRCUMNUCLEAR REGION OF NGC 2110

MARK DURRÉ AND JEREMY MOULD
Centre for Astrophysics and Supercomputing, Swinburne University of Technology,
PO Box 218, Hawthorn, Victoria 3122, Australia

ABSTRACT

High-resolution observations in the near infrared show star clusters around the active galactic nucleus (AGN) of the Seyfert 1 NGC2110, along with a 90 x 35 pc bar of shocked gas material around its nucleus. These are seen for the first time in our imaging and gas kinematics of the central 100pc with the Keck OSIRIS instrument with adaptive optics. Each of these clusters is 2-3 times brighter than the Arches cluster close to the centre of the Milky Way. The core star formation rate (SFR) is 0.3 $M_\odot$ yr$^{-1}$. The photoionized gas (He I) dynamics imply an enclosed mass of 3-4x10$^8$ $M_\odot$.

These observations demonstrate the physical linkage between AGN feedback, which triggers star formation in massive clusters, and the resulting stellar (and SNe) winds, which cause the observed [Fe II] emission and feed the black hole.

## 1. INTRODUCTION

Active galactic nuclei (AGN) are credited as complex agents of feedback, suppressing (Puchwein & Springel 2013), but also enhancing star formation (Mirabel et al. 1999; Mould et al. 2000). Nearby galaxies hint that these two processes are not mutually exclusive, but are closely coupled (Floyd et al. 2013; Rosario et al. 2010). To resolve the physical picture of the star formation-AGN coupling, active nuclei must be put under the magnifier in all the detail that laser guide star adaptive optics (LGSAO) on our largest telescopes can muster.

Eddington (1926) recognized feedback as the primary agent resisting the formation of ever-larger stars. Similarly, theory points to AGN feedback as the controller of galaxy growth, and calls for higher resolution to be brought to bear on "sub-grid physics" that is more complex than the stellar case. Observation is responding to the call with adaptive optics imaging of galactic nuclei. The nuclear starburst connection has been examined by several groups in the infrared with LGSAO (e.g. Bedregal et al. (2009); Brandl et al. (2012); Davies et al. (2007)); they find a variety of formation mechanisms and BH fuelling scenarios.

Our sample for study with the highest resolution is the set of elliptical and S0 galaxies with radio AGN within 40 Mpc. We begin with early type galaxies because their fuelling and stellar population is likely to be the simplest; their stored gas is smaller and there is a minimum of nuclear obscuration. We choose our local volume to correspond to parsec resolution of the circumnuclear gas surrounding these AGN with the best resolution spectrographic integral field unit (IFU). Our sample is based on the Brown et al. (2011) early type radio galaxy catalog, filtered by Palomar Triplespec IR spectroscopy (Mould et al. 2012), selecting those objects that show emission lines, to enable gas dynamics to be observed. This sample is volume complete in the part of the sphere accessible from Hawaii (Keck).

Our first results on the infrared emission line sample are for the galaxy NGC 2110. It is a SAB0$^-$ galaxy (i.e. weakly barred lenticular) (de Vaucouleurs et al. 1991) with Seyfert 1i activity (Véron-Cetty & Véron 2006) – this indicates "normal" Seyfert 2 activity, but with a highly reddened Pa β broad-line region. Others (e.g. Ferruit et al. 2004; Moran et al. 2007; Rosario, et al. 2010; Storchi-Bergmann et al. 1999) state it has Seyfert 2 activity. It has a heliocentric recession velocity of 2335(±20) km s$^{-1}$ (Gallimore et al. 1999), at a distance of 35.6(±1.8) Mpc derived from the Tully-Fisher relationship (Theureau et al. 2007), equivalent to 172 pc arcsec$^{-1}$, and a distance modulus of 32.76(±0.11) mag. (values from the NASA/IPAC Extragalactic Database – NED[1]). It has a radio power of 3.9 × 10$^{22}$ W Hz$^{-1}$ at 1.4 GHz (Ulvestad & Wilson 1989) and X-ray (2-10 keV) power (Nandra et al. 2007) of 2.6 x 10$^{35}$ W. Evans et al. (2006) also find X-ray emission 4" north of the nucleus, co-incident with the edge of the radio jet and the [O III] emission found with Hubble (Mulchaey et al. 1994). The nuclear radio emission is variable (Mundell et al. 2009) (~38% decline in nuclear flux density over seven years); however, Glass (2004) finds minimal near IR variability. With NASA's IRTF, Riffel, Rodríguez-Ardila, & Pastoriza (2006) measured continuum fluxes in the range 1.9 to 2.8 x 10$^{-14}$ erg cm$^{-2}$ s$^{-1}$; a 0.8″ x 15″ slit was employed in one arcsec seeing, giving a spectral resolution of 360 km s$^{-1}$.

Müller, Storchi-Bergmann, & Nagar (2012) and Schnorr-Müller et al. (2013) presented kinematics at moderate resolution; the central H α emission in the Gemini GMOS field displays a rotation pattern, although deviations from simple rotation were seen. Subtraction of a kinematic model of circular rotation revealed two bipolar outflows with velocities of about 100 km s$^{-1}$. They estimated a cold gas inflow rate of 2.2 × 10$^{-2}$ $M_\odot$ yr$^{-1}$ and a warm gas outflow rate of 0.9 M$_\odot$ yr$^{-1}$. Their spatial resolution on the Gemini South telescope was about 100 pc.

---

[1] `http://ned.ipac.caltech.edu/`

Figure 1a shows the combined VLA 6cm image (green contour) from Ulvestad & Wilson (1989), overlaid on the Hubble Space Telescope optical broadband image, which displays a delicate spiral structure (Malkan, Gorjian, & Tam 1998). The blue overplotted boxes represent the OSIRIS narrow-band (solid lines) and wide-band (dotted lines) filter fields of view (FOV) for 0.035 arcsec per pixel plate scale. In this and all subsequent images, the orientation is north up and east to the right. We observed this object with the Triplespec infrared spectrograph at the Mt. Palomar 200" telescope (Mould, et al. 2012) with a 1 arcsec wide longslit, oriented east-west, and spectral resolution R = 2600. The resulting spectrum is plotted (reduced to rest frame) in Figure 1b, showing prominent emission lines of He I (1083.03 nm) and [Fe II] (1256.68nm and 1643.55 nm); the continuum flux values are in line with those of Riffel, et al. (2006). Other lines detected include Pa β, Br γ and H$_2$ in emission, plus the CO bandheads at 2290 and 2322 nm in absorption. The Pa β line showed a dispersion of about 210 km s$^{-1}$, with no indication of Seyfert 1-type broadening.

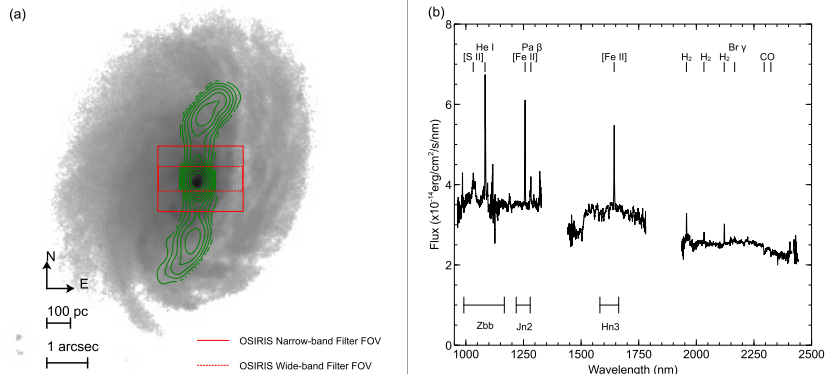

**Figure 1** – (a) HST image from WFPC2 with the F606W filter @ 0.045 arcsec pixel$^{-1}$, is re-oriented and enhanced from the original. The overlayed contours are the VLA 6cm emission, 0.3' x 0.3' @ 0.08 arcsec pixel$^{-1}$. The contours are scaled and re-oriented so east is to the right. The OSIRIS narrow and wide band FOVs are overplotted. (b) Mt. Palomar Triplespec near IR spectrum. The Keck OSIRIS filter bands used are shown (Zbb, Jn2, Hn3).

## 2. OBSERVATIONS AND DATA REDUCTION

We observed the nucleus of NGC 2110 with the OH Suppressing InfraRed Imaging Spectrograph OSIRIS (Larkin et al. 2006) with the Keck I Laser Guide Star Adaptive Optics (LGSAO) system (van Dam et al. 2006; Wizinowich et al. 2006). OSIRIS is a lenslet array IFU spectrograph with a 2048 × 2048 Hawaii-2 detector and spectral resolution R ~ 3600.

We performed observations on this object on 2012 November 1st and 2nd. Each data set (see Table 1) consists of two 900-second object frames, with an intervening 900-second sky frame in the standard object-sky-object sequence. In all observations, the 0.035 arcsec per pixel plate scale was used (giving a scale of 6 pc per pixel). The spectral resolution of OSIRIS is about R = 3800 for this plate scale, but this varies somewhat over the field (Larkin, et al. 2006).

The full width at half-maximum (FWHM) of the tip-tilt stars was approximately two pixels. This gives us a resolution of 70 mas or about 12 pc at 36 Mpc. Seeing ranged from 0.8 to 1.2 arcsec. The observing procedure was to acquire the tip-tilt star (which was in this case the bright core of the galaxy) at position angle zero of OSIRIS to position the object within the unvignetted field-of-view of the LGSAO system. As the radio jet is aligned roughly N-S, we assumed that would be the axis of symmetry of the torus, so the orientation was left in the default position, i.e. the long axis of the field of view was aligned east-west. We concentrated on the strong emission lines of He I and [Fe II] to give the best signal-to-noise ratio (S/N) within the observing time constraints. Alternative emission lines (e.g. Br γ, H$_2$ or Pa β) were not observed; data for these will be obtained either from archived IFU data or future observations.

**Table 1**
Observations

| Filter | Central wavelength (nm) | Range and bandwidth (nm) | Spectral scale (nm/pixel) | Target spectral line | Data sets |
|---|---|---|---|---|---|
| Jn2 | 1260 | 1228-1289 61 | 0.15 | [Fe II] 1257nm | 2 |
| Hn3 | 1635 | 1594-1676 82 | 0.2 | [Fe II] 1644nm | 2 |
| Zbb | 1090 | 999-1176 177 | 0.12 | He I 1083nm | 1 |

Each object frame was reduced using the OSIRIS data reduction pipeline (DRP), using the sky frames for background subtraction. The DRP uses "Rectification Matrices" which incorporate all the flat fielding, bias and dark frames in one, along with mapping each pixel to a lenslet/spectral element. These matrices are retrieved from the Keck repository; for our data, these were dated June 2012. After basic data reduction, the resulting data cube spectra still showed large OH skylines, as the sky background was rapidly changing. The "Scaled Sky Subtract" DRP module (Davies 2007) was used to suppress these skylines.

The resulting data cubes have fields of view of 2.31 x 1.785 arcsec (66 x 51 pixels) and 2.275 x 0.665 arcsec (65 x 19 pixels) for the narrow and wide band filters, respectively. Post data reduction, the FITS file imager and manipulating package QFITSVIEW (Ott 2011) with the underlying DPUSER language was used for analysis. The data cubes were centered and trimmed for display purposes, resulting in images with about 2000 and 500 spectra for narrow and wide band respectively. The resulting data cube values showed some spikes and geometric artifacts that are products of the DRP. These are filtered out by substituting any value that is more than 5 median deviations from the median cube value, with the spatial boxcar median value around the offending pixel.

Telluric correction was applied using observations on A0 stars before and after each target visit, on both nights. These observations were also reduced by the DRP and had the artifacts removed, as before. The DRP then removes the hydrogen absorption lines and divides by an A0V blackbody, then corrects the data cube by this spectrum.

Gas velocity and dispersion mappings, discussed below, are computed using the VELMAP and EVALVELMAP routines within QFitsView. We did not reduce the spectra to rest frame, as there is some uncertainty in the OSIRIS wavelength calibration, and the target lines have good S/N, making identification unproblematic. To increase the S/N ratio for the emission line and kinematic maps, the "Weighted Voronoi Tessellation" (Cappellari & Copin 2003) was used. This aggregates spatial pixels in a region to achieve a common S/N ratio. This procedure requires a noise map, which was taken from the reduced sky background cubes, suitably smoothed. We set the velocity zero point to be the average velocity over the whole map.

The telluric observations were also used as flux calibration sources. The data cubes were reduced as before, and photometry was performed using a suitable aperture. Usually, the AO loops were left open, to prevent the CCD from saturating; this meant that the stellar image could be over 10 pixels wide. The aperture width was chosen for each image where the S/N ratio was at a maximum, i.e. adding any extra width to the aperture just increased noise at the expense of signal. Comparing the flux counts with the published 2MASS infrared magnitudes (and corresponding fluxes) of the stars gives the following calibration constants for each Keck filter band. The zero-point magnitudes were computed by minimizing the differences between the published and observed magnitudes over all observations for that filter. The flux density to data number ratio was taken from the Spitzer Science Center magnitude to flux density converter (Spitzer Science Center 2010). To convert from the narrow-band to broad-band fluxes, we simply multiplied the flux by the ratio of the bandwidth, e.g. for Jn2 to Jbb, the multiplier was 3.8.

To compute the Z band zero-point value, the star was assumed to have the same flux density as in J band. For the flux to DN ratio, the fluxes for the Johnson R, I, J, H and K bands for magnitude 22.7 were interpolated by a power law to obtain the Z band flux. Table 2 gives the results.

**Table 2**
Flux Calibration Data

| Filter | Bandwidth (nm) | Zero-point (mag) | Flux to DN ratio ($\times 10^{-18}$ erg cm$^{-2}$ s$^{-1}$ nm$^{-1}$ per DN s$^{-1}$ nm$^{-1}$) |
|---|---|---|---|
| Zbb | 177 | 22.7 | 3.30 |
| Jbb | 236 | 23.5 | 1.23 |
| Hbb | 330 | 23.7 | 0.417 |
| Kbb | 416 | 22.7 | 0.338 |

The instrumental dispersion was measured using the OH skylines from the off-object data cubes; these lines are assumed to have very narrow width. The average σ value obtained over all filters used was about 40 km s$^{-1}$. This is a lower value than the 2 pixel spectral scale of about 70 km s$^{-1}$, which will be used later.

## 3. RESULTS

### 3.1. Circumnuclear Young Star Clusters

In contrast to other observations, ours are confined to the nuclear region of this galaxy, presumably encompassing the dusty torus and obscured AGN/BLR. Star formation in the circumnuclear region of NGC 2110 was best resolved in the J-band continuum. This has the appearance, as shown in Figure 2a, of four star clusters (labeled "A" to "D") in a semi-circular ring, having a major axis of about 90 pc. In this and following maps, the location of the AGN is identified by "X" near cluster "B"; see below for discussion of the identification. The clusters are best resolved in one of the J band data sets, the others and the H-band data sets show them less clearly; seeing limitations are probably responsible.

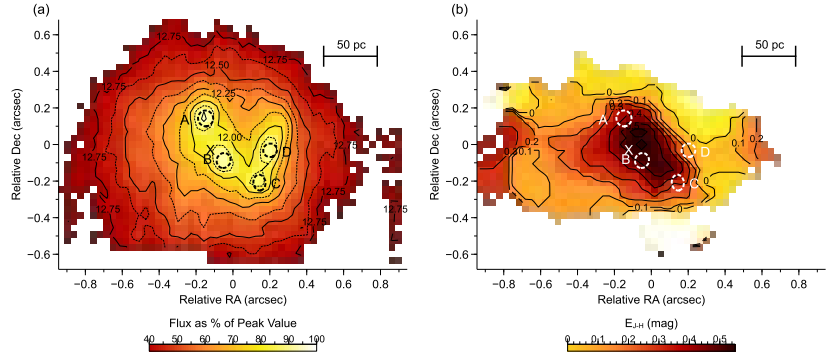

**Figure 2** – (a) The OSIRIS Jn2-band (1228-1289 nm) continuum of the nuclear region. The contours are in apparent magnitude per arcsec$^2$. The shading represents the flux as a percentage of the peak value. The overplotted ellipses are the best fit to the deconvolved flux FWHM. (b) Extinction Map E(J-H).

**Table 3**
**Cluster Luminosities and Sizes**

| Cluster | Size (FWHM) (pc) | Flux$_J$ (x10$^{-15}$ erg cm$^{-2}$ s$^{-1}$) | Apparent magnitude (m$_J$) | Absolute magnitude (M$_J$) | L/L$_\odot$ (x 10$^8$) |
|---|---|---|---|---|---|
| A | 18x22 | 2.5 | 15.2 | -17.5 | 2.9 |
| B | 26x17 | 2.2 | 15.4 | -17.4 | 2.6 |
| C | 21x20 | 2.0 | 15.5 | -17.3 | 2.3 |
| D | 16x21 | 1.4 | 15.9 | -16.9 | 1.6 |

They are not visible in the Z-band images; the smaller FOV misses some of them, as well as there being reduced AO performance at shorter wavelengths.

To measure the luminosity of the clusters, we performed the following process. We fitted a point-spread function to a Jn2 band telluric star observation on that night (with loops closed), then performed a Lucy-Richardson deconvolution using the IRAF[2] package *stsdas.analysis.restore.lucy*, with 20 iterations. For each cluster in the deconvolved image, we fitted an elliptical Gaussian profile and estimated the data number flux. Since this was the narrow-band measurement, this was scaled to get an estimated total J band count (as described above). The magnitude was then calculated from the above calibration, using the distance modulus given above and the solar J band absolute magnitude of 3.64 (Binney & Merrifield 1998). Table 3 gives the results. These clusters can be compared to the Arches cluster in the center of the Milky Way, with J band luminosity L = 0.6 x 10$^8$ L$_\odot$ (Figer et al. 2002) and the Hercules globular cluster M13, L = 3.6 x 10$^5$ L$_\odot$ (from SIMBAD[3] and NED).

Subtracting the star cluster flux from the original image reveals a rather flat field with some loops and arc segments, but impression is of a ring with a weak central emission. The next strongest source has a flux of 0.9 x 10$^8$ L$_\odot$ - this can be regarded as a northward extension of cluster "D".

### 3.2. *Gas Distribution and Kinematics*

The Jn2, Hn3 and Zbb spectra of the central 0.75 arcsec (encompassing the clusters) are presented in Figure 3; the wavelengths have not been reduced to the rest frame.

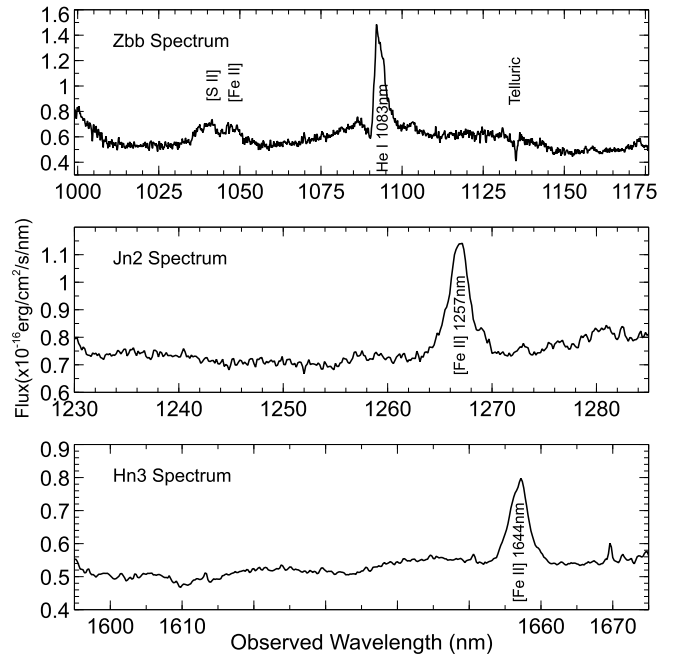

**Figure 3** – OSIRIS spectra of the central 0.75 arcsec (a) Jn2, (b) Hn3 and (c) Zbb.

---

[2] http://iraf.noao.edu/

[3] http://simbad.u-strasbg.fr/simbad/

The region of star formation is also seen in the shocked [Fe II] gas. Figure 4 shows the [Fe II] flux, velocity and dispersion maps for 1644 nm. This provides a slightly clearer picture than the shorter wavelength [Fe II] image (Figure 5), due to somewhat greater extinction at 1257 nm. The flux maps show a roughly elliptical region or broad bar, about 90 x 35 pc in size, which can be interpreted as a dusty cylindrical or toroidal obscuring region around the central black hole (BH). The position angle (PA) of the bar is 135-315°, i.e. at about 45° inclination to the north-south radio jet, where 0° is north and 90° is east. Smoothing the flux with a 3-pixel Gaussian function produces the overplotted contours in green in Figure 4a, showing leading and trailing arms on the bar which, as we will see later, correspond to heavily obscured regions in the HST image and fit in with the overall spiral structure of the inner kiloparsec. We can estimate the inclination of the toroidal structure by fitting an ellipse to the outer flux contour in Figure 4a. This gives a minor to major axis ratio of about 0.7, implying an inclination of about 45˚ (where 0˚ is face-on). This is calculated under the assumption that the disk is thin in comparison to its width; any appreciable scale height will increase the calculated angle.

Note that the maximum [Fe II] flux is not collocated with the clusters, but rather lies between them. This is the shocked inter-cluster medium that is radiating, excited by strong outflows. The AGN does not appear to contribute directly to the observed outflows; these are localized to the clusters and bar/torus and do not seem to relate to the existing optical, radio or X-ray data. The outflows are driven by star formation, evolved stellar winds and SN shocks. The presence of dust suggests NGC 2110 would repay investigation with ALMA of its molecular gas phase at this resolution.

The velocity maps show a rotational structure; this will be explored more fully in discussions on the He I observations, below. The dispersion map does not show much structure, but has σ values in the range 150 to 300 km s$^{-1}$, consistent with Seyfert narrow line emission values.

Moran, et al. (2007) gives a BH mass $2 \times 10^8$ M$_\odot$, based on the M-σ relationship and the measured dispersion of 220 km s$^{-1}$. As we see later, this is also supported by our observations of the He I and [Fe II] spectral lines. The clusters seem embedded in the general gas velocity field. We are deducing the velocities for the clusters from the gas rather than from stellar light, as no stellar absorption features that could be used to determine the cluster velocities were seen.

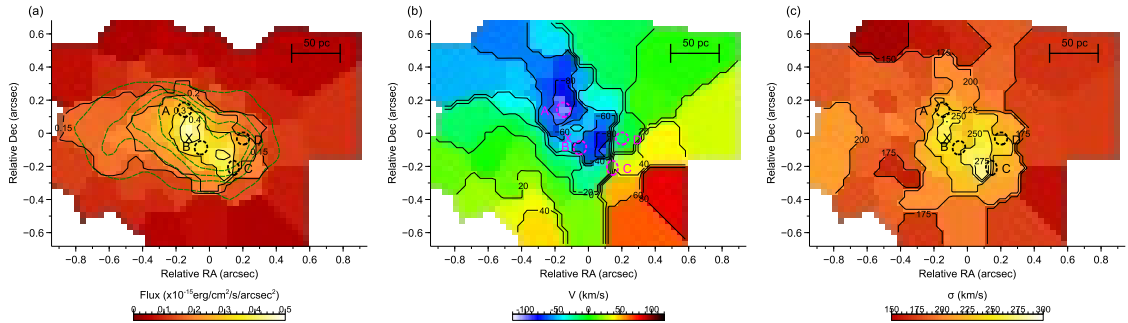

**Figure 4** – OSIRIS [Fe II] 1644nm (a) flux. (The overplotted contours in green are the flux values smoothed by a 3-pixel Gaussian function to delineate the structure in arbitrary units), (b) velocity and (c) dispersion.

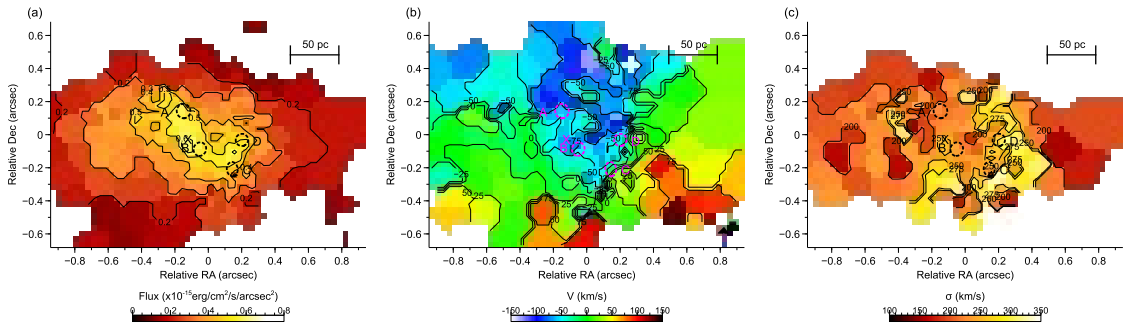

**Figure 5** – [Fe II] 1257nm (a) flux, (b) velocity, (c) dispersion.

The location of the nucleus of the galaxy (and presumably the AGN) in the Jn2 and Hn3 maps is an interesting problem, as there are no readily defined features to map onto (say) the HST image. To match the images, we did the following procedure. We smoothed the [Fe II] flux map with a 3 pixel Gaussian function - this clearly shows the bar and 2 trailing arms. We took a radial profile of the HST image from the brightest pixel and constructed a circularly averaged image, which was used to numerically divide the original image. This enhanced the spiral and jet features. Matching the brightest pixel in both images, the spiral features in the optical are seen to merge with the IR emission line flux, with the arms of the bar in the [Fe II] image tracing the dusty lanes in the HST image (Figure 8). We can thus identify the AGN/galactic nucleus location (marked with an "X" on all the [Fe II] the maps). Note that the orientation of the jet in the HST image is orthogonal to the bar in the extinction image, enhancing the identification.

Riffel et al. (2013) found a correlation between the stellar and [Fe II] gas dispersions. The average dispersion of the [Fe II] 1257nm spectral line is 225 km s$^{-1}$. Subtracting the instrumental dispersion in quadrature from this value, the correlation gives a stellar dispersion of 148(±45) km s$^{-1}$.

### 3.3. Extinction by Dust

Extinction values can be derived from the [Fe II] 1644/1257 nm ratio: these lines share the same atomic upper level ($a^4D \to a^6D$ and $a^4D \to a^4F$, respectively), and thus their theoretical ratio is a function of the frequencies and probabilities of the transition and is independent of the plasma conditions. From National Institute of Standards and Technology (National Institute of Standards and Technology 1995), the transition probabilities (Einstein coefficients) $A_{ki}$ are 4.74 and 6.0 x 10$^{-3}$ s$^{-1}$ respectively. Thus the emitted flux ratio (which is inversely proportional to the wavelength) is $F_{1257}/F_{1644}$ = 1.0332; the intrinsic magnitude difference is thus 0.035.

The flux value at each pixel can be calibrated from the telluric star observations, taking the average spectral data numbers of 10 wavelength pixels around the central wavelength. The extinction is then estimated for each pixel by $E_{J-H} = m_{1257} - m_{1644} + 0.035$, where $m_\lambda$ is the measured magnitude at wavelength λ. The resulting map is median smoothed for each pixel by 3 x 3 pixels, as shown in Figure 2b. This presents a broad central region, encompassing the clusters, obscured to about 0.5 magnitudes. The standard deviation of each pixel from the smoothed value is about 0.3 magnitudes. Over the central 1 arcsec, we get an average extinction of about 0.1 mag. Uncertainties in the flux ratios around the periphery produce unphysical negative values that should be ignored. As a further check on our OSIRIS calibration and extinction calculations, we compare the ratio of the two spectral line fluxes in both the OSIRIS and the Triplespec data; these are the same to within about 3%.

### 3.4. He I and the Nucleus

The atomic structure of He I makes it a unique astrophysical tool. The 2s level of the triplet state is metastable with a lifetime of 2 hours, and it acts as a second ground state. The 2s level lies 19.75 eV above the ground state, and so it is populated by recombination. It is depopulated predominantly by collisions in photoionized gas. Although helium is abundant in astronomical gas, He I* is rare. One of every ~175,000 He+ ions is He I* in a typical plasma (Clegg 1987). Our observations are of that state decaying by the λ=1083.0 nm (2s→2p) line.

The broadband Z-band image and He I flux, velocity and dispersion maps (Figure 6) are different from the J band and [Fe II] maps. Both the morphology (more centrally concentrated) and the velocity dispersion (roughly 1.5 times) compared to [Fe II] support the conclusion that what we are seeing at 1083 nm is closer to the photoionizing source. We note the northeast extension of the bright central core is also seen on the HST image (Figure 1a). The velocity map is also oriented in the same direction and sense as the [Fe II] 1644 nm map.

On examination of the spectrum over the whole of the central region (Figure 7a) we can see a P Cygni type profile, indicating an absorbing outflow in the line of sight. Fitting an emission Gaussian curve to the right half of the spectral line and subtracting it from the whole spectrum leaves a residual, which can then be fitted by an absorption line. This shows an emission line of about 655 km s$^{-1}$ dispersion, with a blue-shifted absorption line (indicative of an expanding shell of gas) of about 520 km s$^{-1}$ blue-ward of the emission peak with a dispersion of about 310 km s$^{-1}$. A similar outflow was also measured in He I in NGC4151 by Iserlohe et al. (2013). Taking a cut across the velocity map between the velocity extreme value positions produces the profile as shown in Figure 7b. We assume a simple Keplerian rotation as given by M=V$^2$R/G. The enclosed mass, from the peak-to-peak positions (56 pc) and velocities (356 ± 33 km s$^{-1}$) and the assumed 45° inclination, is calculated as 4.2(+1.2,-0.6) x 10$^8$ M$_\odot$. The error bars are derived from the VELMAP calculation, described above. Fitting a Plummer model to this profile (Riffel et al. 2008; Storchi-Bergmann et al. 1996) gives 3.2 x 10$^8$ M$_\odot$. These values are in line with the M-σ relationship (Moran, et al. 2007). Using the Graham et al. (2011) updated relationship for elliptical galaxies and the estimated stellar dispersion calculated above, we derive a somewhat lower value of 4.9 (+14, - 3) x 10$^7$ M$_\odot$. A similar calculation using the [Fe II] 1644 nm velocity field also gives a lower value, probably due to the confusing outflows from the clusters. Krajnovic, Sharp, & Thatte (2007) have suggested that [Fe II] is not a good tracer of the central potential and is not suitable for determination of BH mass. This simplistic model will over-estimate the BH mass, as there will be non-negligible stellar and gas mass within that radius. We note, however, that for a stellar dispersion (derived above) of 148 km s$^{-1}$ and a BH mass of 2 x 10$^8$ M$_\odot$, the sphere of influence, calculated as r$_h$=GM/σ$^2$ is about 40 pc. i.e. about the scale of the observed nuclear region. We will be presenting more detailed modeling of the rotation and inflow/outflow fields in a future paper, which will also incorporate archival data on K-band stellar data.

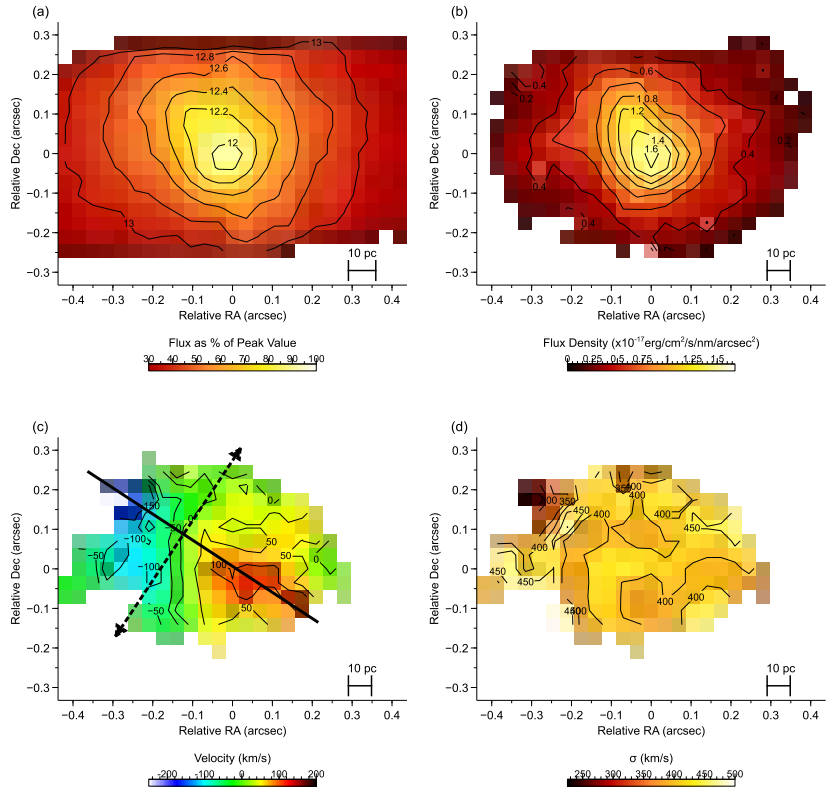

**Figure 6** – (a) The Zbb-band (999-1176 nm) continuum of the nuclear region, (b) the He I emission-line 1083nm flux, (c) He I 1083nm velocity and (d) He I 1083nm dispersion.

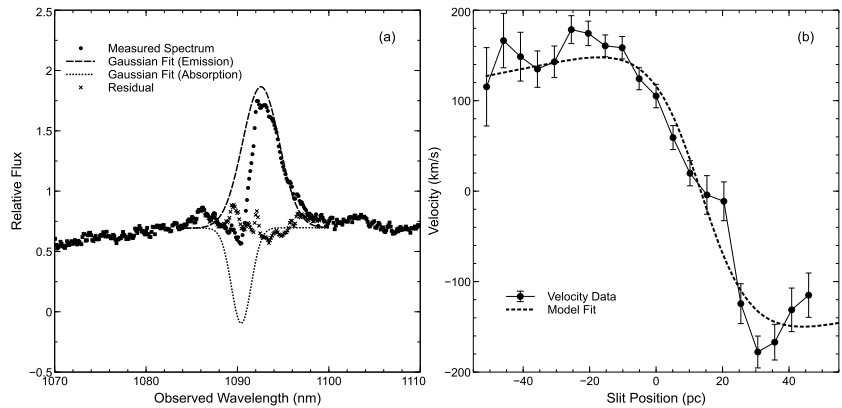

**Figure 7** – (a) He I spectral line with double Gaussian fit in emission and absorption. (b) He I velocity profile and Plummer model fit.

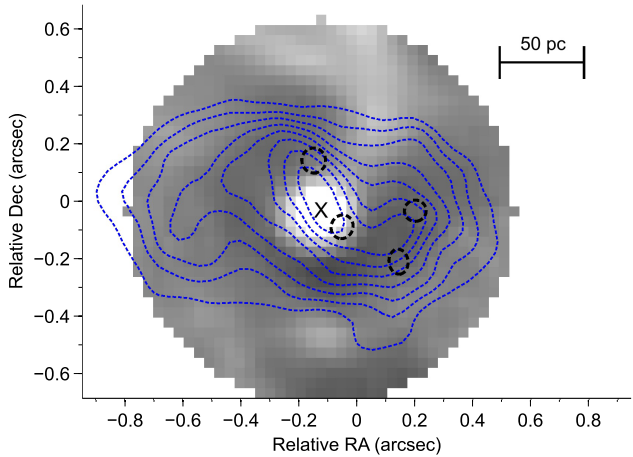

**Figure 8** – Enhanced HST F606W image and smoothed [Fe II] flux contours (arbitrary units), locating the AGN/galactic nucleus.

## 4. DISCUSSION

### 4.1. Cluster Formation and Ages

The Arches cluster near the center of the Milky Way has a mass of $2(\pm 0.6) \times 10^4$ $M_\odot$ (Espinoza, Selman, & Melnick 2009), an age of 2–3 Myr, a luminosity $L = 0.6 \times 10^8$ $L_\odot$ and a small size of 0.2pc; the intrinsic extinction is not substantial (Figer, et al. 2002). Assuming a similar age and mass-to-light ratio, we have about 15.7 Arches masses, i.e. $3.1 \times 10^5$ $M_\odot$ in the circumnuclear region, equivalent to a SFR of about 6.3 Arches clusters per Myr, i.e. about 0.12 $M_\odot$/yr.

Although we do not have a Brγ map like the previous figures, we can use the value of $2.5 \times 10^{-15}$ ergs cm$^{-2}$ s$^{-1}$ from Palomar Triplespec (Mould, et al. 2012), giving a Brγ luminosity of $3.2 \times 10^{38}$ erg s$^{-1}$. The star-forming rate equation of Kennicutt, Tamblyn, & Congdon (1994) is SFR ($M_\odot$ yr$^{-1}$) = $8.2 \times 10^{-40}$ L(Br γ) (erg s$^{-1}$). From this equation, we get 0.26 $M_\odot$ yr$^{-1}$, i.e. there is no difficulty forming these clusters in a few million years.

The dominant component of the infrared emission line spectrum resembles the "Homunculus" nebula around the Luminous Blue Variable η Car (Smith 2002). We can compare the fluxes for the [Fe II] 1257 nm emission line between the two objects; the surface brightness of the brightest [Fe II] flux pixel is $3.8 \times 10^{-16}$ erg cm$^{-2}$ s$^{-1}$ arcsec$^{-2}$; the fluxes from Smith (2002) have a range of 1.2 to $32 \times 10^{-16}$ erg cm$^{-2}$ s$^{-1}$ arcsec$^{-2}$. We can infer the presence of massive stars in the ring around the NGC 2110 nucleus and the consequent violent mass loss associated with their advanced evolution. Hur, Sung, & Bessell (2012) estimated the age of the clusters Tr 14 and Tr 16 in η Car at 1-3 Myr; we adopt the same age for the NGC2110 clusters, based on the similarity of the spectra.

At the radius of the star cluster ring we would expect about a speed of about 150 sin $i$ km s$^{-1}$ and a period of 1.2 Myr for a circular orbit, which means that the clusters are just a couple of dynamical times old and still to reach equilibrium with the BH. The low velocity amplitude would reflect that; furthermore we are observing shocked gas velocities towards the outside of a molecular obscuring region, rather than star clusters in a circular orbit.

Renaud et al. (2013), in their galactic hydrodynamical simulations, noted that star cluster formation close around the central BH seems to be controlled by tidal shear and resonances with spiral arms, and can produce matter condensations at ~ 40pc at the edge of a nuclear disk ("bead on a string"). Just such a mechanism can be postulated for the formation of the observed clusters. Mould, et al. (2000) also observe star formation associated with the radio jet of Cen A and tentatively proposed photoionizing shocks from the jet as the physical mechanism.

In our speculative scenario, the arms are the gas inflow feeders to the bar. The AGN polar bi-conal jets (as seen in HST data) provide confinement and compression to the turbulent gassy/dusty torus, which triggers the formation of the observed massive clusters, which in turn provide the stellar winds (and occasional SN shocks) to cause the observed [Fe II] emission, as shown by the present data. This interaction occurs if the jets intersect the torus; Müller-Sánchez et al. (2011) shows that this commonly occurs and Rosario, et al. (2010) propose this specifically for NGC2110. Some of these cluster outflows feed the AGN, which helps to overcome the problem whereby gas infalling from a large distance has too much angular momentum to impinge directly onto the BH/accretion disk. Further support for this will require more detailed analysis of the flow structure.

### 4.2. Extinction and Gas Mass Estimation

According to Smith (2002), η Car's 1.644/1.257 flux ratio varies from 1.0 to 1.45 at 3 different locations in the nebula. This is the same as our observations (0 to 0.5 mag. $E_{J-H}$). The expectation for normal reddening is $E_{J-H}/E_{B-V} = 0.37$ (Mathis 1990). This gives $E_{B-V}$ up to 1.4 mag., which is consistent with the measurements from Storchi-Bergmann, et al. (1999). With an $R_V = 3.1$ reddening law (Schlafly & Finkbeiner 2011) and a foreground Milky Way (MW) galactic extinction $A_V$ of 1.02 (NED), we obtain a total extinction value $A_V$ up to 3.2 mag. This is consistent with the value given by Alonso-Herrero et al. (1998), whose observations can be described by an evolved stellar population reddened by a foreground extinction of $A_V \approx 2 - 4$ mag, also consistent with the values derived from Hubble images (Mulchaey, et al. 1994). The MW galactic extinction value, however, has some uncertainty in it; the map of this low galactic location (l = 213°, b= -16°), from Burstein & Heiles (1982), shows some patchiness. While we would not argue that the galactic reddening is zero, we can regard the value as an upper limit. Our Palomar Triplespec spectrum (Figure 1b) averages over the whole nuclear region with 1 arcsec resolution (about the same size as our high resolution maps), and shows the same extinction value (about 0.1 mag) as these observations. Our measurements suggest that the massive star formation in NGC 2110's circumnuclear region is up to four times higher and at a similar level of extinction to that of the fan region of the η Car nebula.

Archival HST images of the nucleus, with the Holtzman et al. (1995) calibration, yield V–I ≈ 1.8 mag. for the central tenth of an arcsec. Given the reddening calculated above, this suggests an intrinsic stellar population color of zero. The star clusters visible to OSIRIS in the NIR are not seen in Figure 1a, because of HST's lesser resolution, the archival frame's short exposure and over 3 magnitudes of extinction.

The gas-to-extinction ratio, $N_H/A_V$ varies from 1.8 (Predehl & Schmitt 1995) to $2.2 \times 10^{21}$ cm$^{-2}$ (Ryter 1996). Taking all pixels in the central region that have a $E_{J-H} > 0.1$ mag (328 pixels), we derive an average extinction $A_V$ (less galactic extinction) of 1.4 mag. Using an average atomic weight of 1.4 and assuming rotational symmetry, the calculated total gas mass is $7.5 \times 10^5$ M$_\odot$.

There are interesting features of the emission line fluxes and kinematics. For the He I emission, the rotational center does not seem to be located at the maximum flux position. This could be due either to obscuration towards the observer or from shielding of the gas from the photoionizing source. This displacement is about 20pc to the north-east for the He I line; for the [Fe II] 1644 nm line the gas is excited by shocks from the star clusters that seem to be on one side of the central BH. This displacement is also found by Storchi-Bergmann, et al. (1999). Also, the orientation of the flux and velocity maps is not orthogonal to the radio jet, as would be expected in a simple model of the torus. There is no strong reason it should be; the radio jet also shows evidence of precession.

Rosenberg, van der Werf, & Israel (2012) presented a formula for the supernova (SN) rate, based on the measured [Fe II] 1257 nm flux. SN remnant (SNR) shock fronts destroy dust grains by thermal sputtering, which releases the iron into the gas phase where it is singly ionized by the interstellar radiation field. In the extended post-shock region, [Fe II] is excited by electron collisions (Mouri, Kawara, & Taniguchi 2000), making it a strong diagnostic line for tracing shocks. For the pixel with the greatest flux ($4.8 \times 10^{-19}$ erg cm$^{-2}$ s$^{-1}$) and taking each pixel as 36.5 pc$^2$, we get log $\nu_{SNR}$ [yr$^{-1}$ pc$^{-2}$]= -5.87(±0.9), i.e. about 1 SN per pc$^2$ every 750 kyr.

In the AGN Unified Model, the size of the cylindrically symmetric, smooth obscuring torus is usually given as of the order of several parsecs. What we observe is substantially larger (~90 pc). However, recent 3D hydrodynamic simulations (Schartmann et al. 2008; Wada & Norman 2002; Wada, Papadopoulos, & Spaans 2009) suggest highly inhomogeneous and turbulent obscuring structures, with a radius of several tens of parsecs. Hicks et al. (2009) provide observational support for this scenario with their H$_2$ gas data from a sample of 9 local Seyfert 1 galaxies, showing thick, clumpy gas disks with a typical radius of about 30 pc.

## 5. CONCLUSIONS

LGSAO resolution and IFU spectroscopy at the Keck Observatory of the nucleus of NGC 2110 have shown four massive young star clusters (the brightest of which has $L/L_\odot = 2.7 \times 10^8$) embedded in a disk of shocked gas, with an estimate of the enclosed mass of 3.2 to $4.2 \times 10^8$ M$_\odot$. Though NGC 2110 is an S0, its nuclear region is being fuelled sufficiently on million year timescales to sustain a star formation rate of order 0.3 M$_\odot$ yr$^{-1}$, in line with the cluster formation rate we see.

The process that terminates galaxy growth by accretion is called feedback. Black holes are integral to that process, found in all massive galaxies, and, when feedback is active, are AGN. Activity classically embodies radio jets that can entrain and drive material clean out of galactic halos. But nuclear feedback can also be mediated as star formation, in which massive young stars eject a major fraction of their mass in powerful winds. In this galaxy we are witnessing one of the modes of AGN feedback at a few parsecs resolution, the formation of massive stars with energetic winds.

## 6. ACKNOWLEDGEMENTS

The data presented herein were obtained at the W.M. Keck Observatory, which is operated as a scientific partnership among the California Institute of Technology, the University of California and the National Aeronautics and Space Administration. The Observatory was made possible by the generous financial support of the W.M. Keck Foundation. The authors wish to recognize and acknowledge the very significant cultural role and reverence that the summit of Mauna Kea has always had within the indigenous Hawaiian community. We are most fortunate to have the opportunity to conduct observations from this mountain.

We would like to thank our Triplespec colleagues at Caltech for their ongoing support of our survey. The Jet Propulsion Laboratory, California Institute of Technology operates the NASA/IPAC Extragalactic Database (NED), under contract with the National Aeronautics and Space Administration. The SIMBAD Astronomical Database is operated by CDS, Strasbourg, France. IRAF is distributed by the National Optical Astronomy Observatories, which is operated by the Association of Universities for Research in Astronomy, Inc. (AURA) under cooperative agreement with the National Science Foundation. We also thank the anonymous referee for helpful comments.